\documentclass[4apaper]{aipproc}

\layoutstyle{8x11double}
\newcommand{\Ord}{\mathcal{O}}
\newcommand{\csw}{c_{SW}}
\newcommand{\ev}[1]{\left\langle #1 \right\rangle}

\title{Towards the $N_f=2$ deconfinement transition temperature with $\Ord(a)$ improved Wilson fermions: An update}

\author{Bastian B. Brandt}{address={Institut f\"ur Kernphysik, Johannes Gutenberg-Universit\"at Mainz, Johann-Joachim-Becher-Weg 45, 55128 Mainz, Germany},email={brandt@kph.uni-mainz.de}}
\author{Owe Philipsen}{address={Institut f\"ur Theoretische Physik, Goethe-Universit\"at, Max-von-Laue-Str. 1, 60438 Frankfurt am Main, Germany}}
\author{Hartmut Wittig}{address={Institut f\"ur Kernphysik, Johannes Gutenberg-Universit\"at Mainz, Johann-Joachim-Becher-Weg 45, 55128 Mainz, Germany}}
\author{Lars Zeidlewicz}{address={Institut f\"ur Theoretische Physik, Goethe-Universit\"at, Max-von-Laue-Str. 1, 60438 Frankfurt am Main, Germany}}

\keywords{Lattice QCD, Quark-gluon plasma, Deconfinement, Phase transitions \vspace*{-8cm} \newline \hspace*{.8\textwidth} {\normalsize {\tt MKPH-T-10-38}} \vspace*{7.1cm} \newline}

\begin{abstract}
We give an update on our current project to determine the transition temperature and the order of the deconfinement transition in the chiral limit of two flavour QCD. We use nonperturbatively $\Ord(a)$ improved Wilson fermions of the Sheikholeslami-Wohlert type, employing the efficient deflation accelerated DDHMC algorithm. We start at lattices with $N_t\geq12$ and pion masses below 600 MeV, aiming at chiral and continuum limits with light quarks.
\end{abstract}

\begin{document}

\maketitle

\section{Introduction}

The transition from hadronic matter to the deconfined state of free quarks and gluons is a prime subject of current research in elementary particle physics. From the theoretical point of view two of the most important questions concern the transition temperature and the order of the transition at zero and finite chemical potential. In this context the role of the related chiral symmetry restoration is of special interest. A perturbative treatment of the plasma at the transition temperature is not valid, thus lattice QCD is the preferred tool to study the transition.

Most of the state-of-the-art simulations so far have been performed using staggered
fermions, having the advantage of being numerically cheap compared to
other fermion discretisations. Recent results with 2+1 flavours of dynamical quarks can be found in
refs.~\cite{hotQCD,BW-group}. However, there are conceptual problems concerning the staggered approach to lattice QCD (see e.g. the discussions in \cite{stagg}) and a cross-check of the
staggered results is needed using other fermionic
discretisations. Several groups have already started to perform
simulations with $\Ord(a)$ improved
Wilson fermions of the Sheikholeslami-Wohlert type
\cite{QCDSF,WhotQCD} as well as with maximally twisted mass
\cite{tmft}. All these simulations still suffer from unphysically large pion masses and
lack continuum extrapolations.

In this proceedings article we give an update on the results concerning our study of the $N_f=2$ phase transition, using non-perturbatively $\Ord(a)$-improved Wilson fermions, lattices with $N_t\geq12$ and pion masses lower than 600 MeV. We aim to extract the transition temperature and the order of the transition in the chiral limit, which is still not settled until today. There are two scenarios \cite{scen1,scen2}: In the first scenario, the chiral critical line in the $\{m_{u,d},m_{s},T\}$-parameter space never reaches the $m_{u,d}=0$ axis, while the second one implies the existence of a tricritical point at $m_{u,d}=0$, which extends into the direction of finite chemical potential as a critical line.
With our $N_f=2$ simulation, we therefore address a question which is important for the enlarged phase diagram of the $N_f=2+1$ theory as well as on the phase diagram at finite density.

\section{Setup of the simulations}

\begin{table}[t]
\centering
\begin{tabular}{c|cccccccc}
\hline
scan & Lattice & Block size & $\kappa$ & $\beta$-range & $\tau$ & $\tau_{int}[P]$ & $f_{meas}$ & Statistic \\
\hline
$A$ & $12\times24^3$ & $6^4$ & 0.13595 & $5.270-5.320$ & 2.0 & $\Ord(30)$ & 1 & $\Ord(25000)$ \\
$B$ & $16\times32^3$ & $8^4$ & 0.13650 & $5.400-5.575$ & 2.0 & $\Ord(10)$ & 2 & $\Ord(5000)$ \\
\hline
\end{tabular}
\caption{Run parameters for scans $A$ and $B$. We show the DDHMC block size, the Monte Carlo time $\tau$ of the trajectories, the measurement frequency $f_{meas}$ and the integrated autocorrelation time $\tau_{int}$ of the plaquette $P$.}
\label{tab1}
\end{table}

We employ two degenerate flavours of nonperturbatively $\Ord(a)$ improved Wilson fermions, using the Sheikholeslami-Wohlert lattice Dirac operator \cite{SW-action}
\begin{equation}
 \label{eq-sw-op}
D_{SW} = D_W + \csw \: \frac{i\:a\:\kappa}{4} \: \sigma_{\mu\nu} \: \hat{F}^{\mu\nu} \; .
\end{equation}
Here $D_W$ is the usual Wilson Dirac operator, $\kappa$ is the hopping parameter, $\sigma_{\mu\nu}$ the totally antisymmetric tensor and $\hat{F}^{\mu\nu}$ the 'clover leaf' representation of the gluonic field strength tensor on the lattice. The clover coefficient $\csw$ is tuned with $\beta$, using the interpolation formula from \cite{npcsw}.
The simulations are generated using the deflation accelerated DDHMC algorithm, introduced by L\"uscher \cite{DDHMC}. This algorithm is also used intensively in the context of the CLS effort \cite{CLS_wiki} for simulations at zero temperature \cite{CLS,CLS_POS}.

To scan the temperature, we vary the lattice spacing $a$ via the bare lattice coupling $\beta$, which is connected to the temperature by $T=1/[N_t\:a(\beta)]$. This method enables us to get a fine resolution around the critical temperature, in contrast to the fixed scale approach, and to use the modified Multi-Histogram method as introduced in \cite{ownpos}. The scale is set after the determination of the critical coupling $\beta_c$ by an additional run at $T=0$.

\begin{figure}[t]
\centering
\includegraphics[angle=-90, width=.38\textwidth]{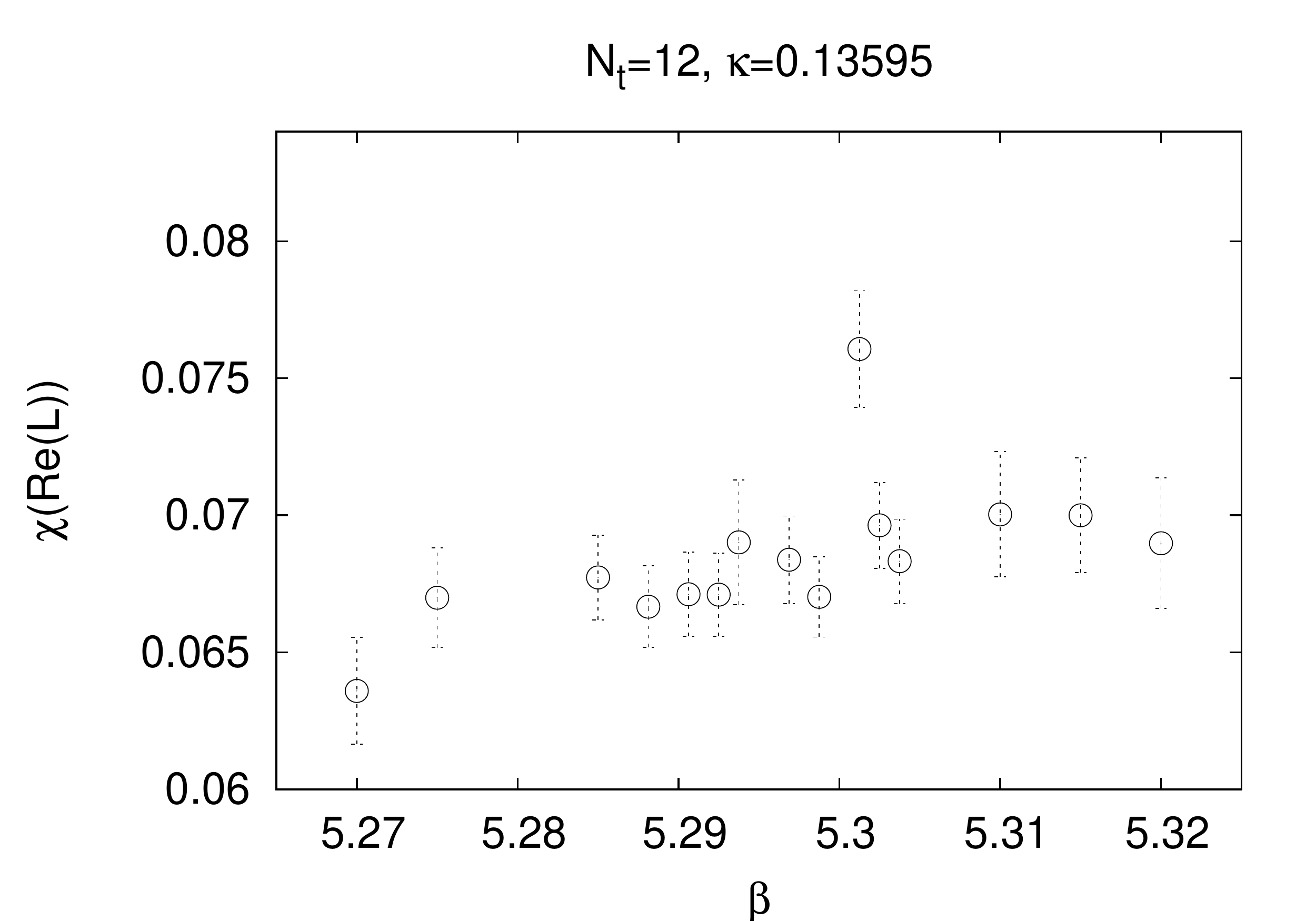}
\caption{Plot of the results for the Polyakov loop suszeptibility of scan $A$.}
\label{fig0}
\end{figure}

To investigate the properties of the finite temperature transition we look at the behaviour of the average plaquette $\ev{P}$, the real part of the Polyakov loop $\textnormal{Re}\left[\ev{L}\right]$ and the chiral condensate $\ev{\bar{\psi}\psi}$. We also find it beneficial to compute the Polyakov loop in an APE-smeared version since the smearing leads to a more pronounced signal for the phase transition, as already observed in \cite{hw-89}.

We define the generalised susceptibilities $\chi(O)$ by
\begin{equation}
 \label{susz}
\chi(O) \equiv N_s^3 \: \left( \ev{O^2} - \ev{O}^2 \right) ,
\end{equation}
where $O$ is any of the observables above and $N_s$ the spatial lattice size. These generalised susceptibilities should show a notable peak at the transition point. In addition, the behaviour of the peak under a change in the spatial volume is governed by the corresponding critical exponents, encoding information about the order of the transition.

\section{Simulation results}
\label{results}

\begin{figure}[t]
\centering
\begin{minipage}{.45\textwidth}
\centering
\includegraphics[angle=-90, width=.85\textwidth]{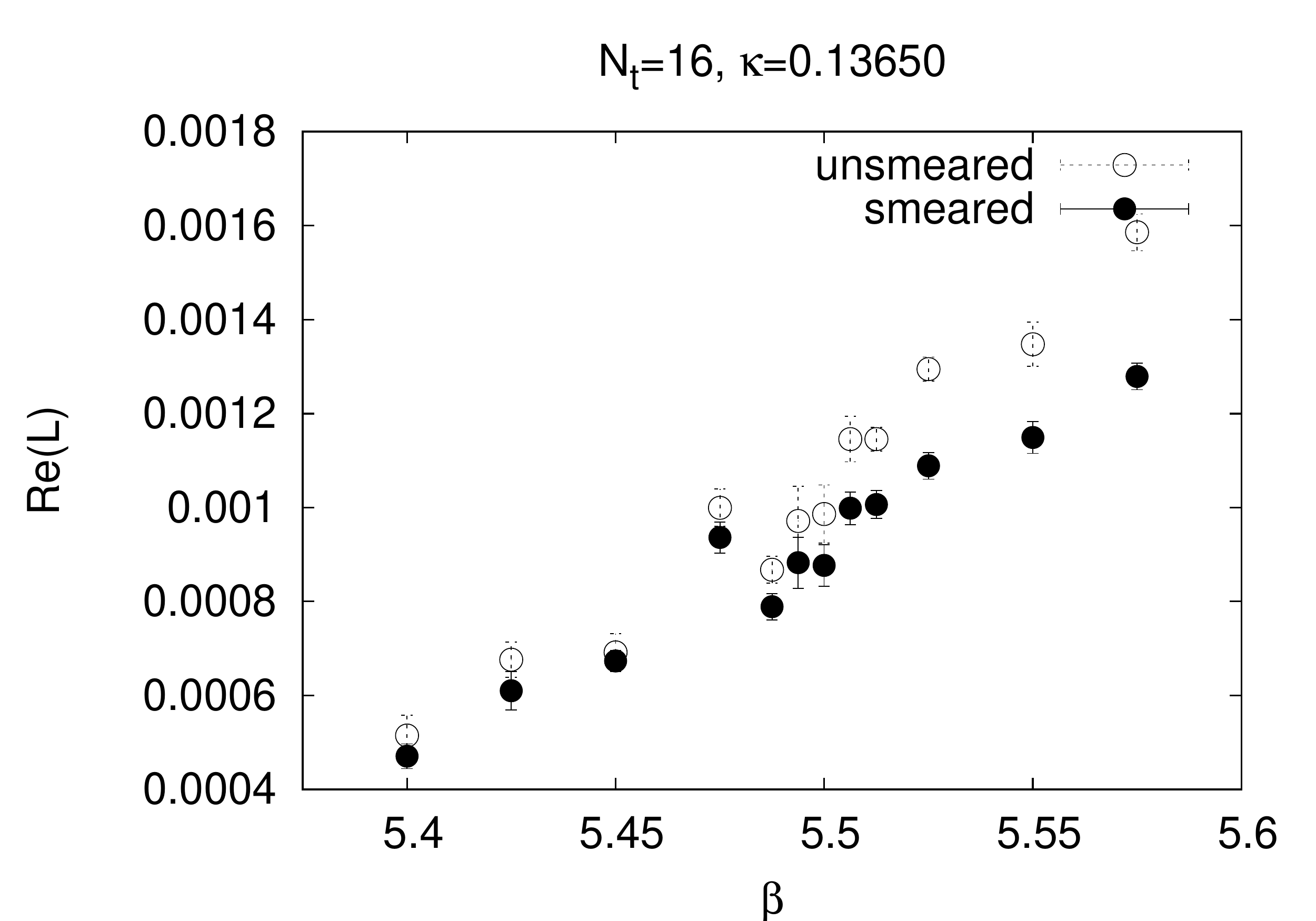}
\end{minipage}
\begin{minipage}{.45\textwidth}
\centering
\includegraphics[angle=-90, width=.85\textwidth]{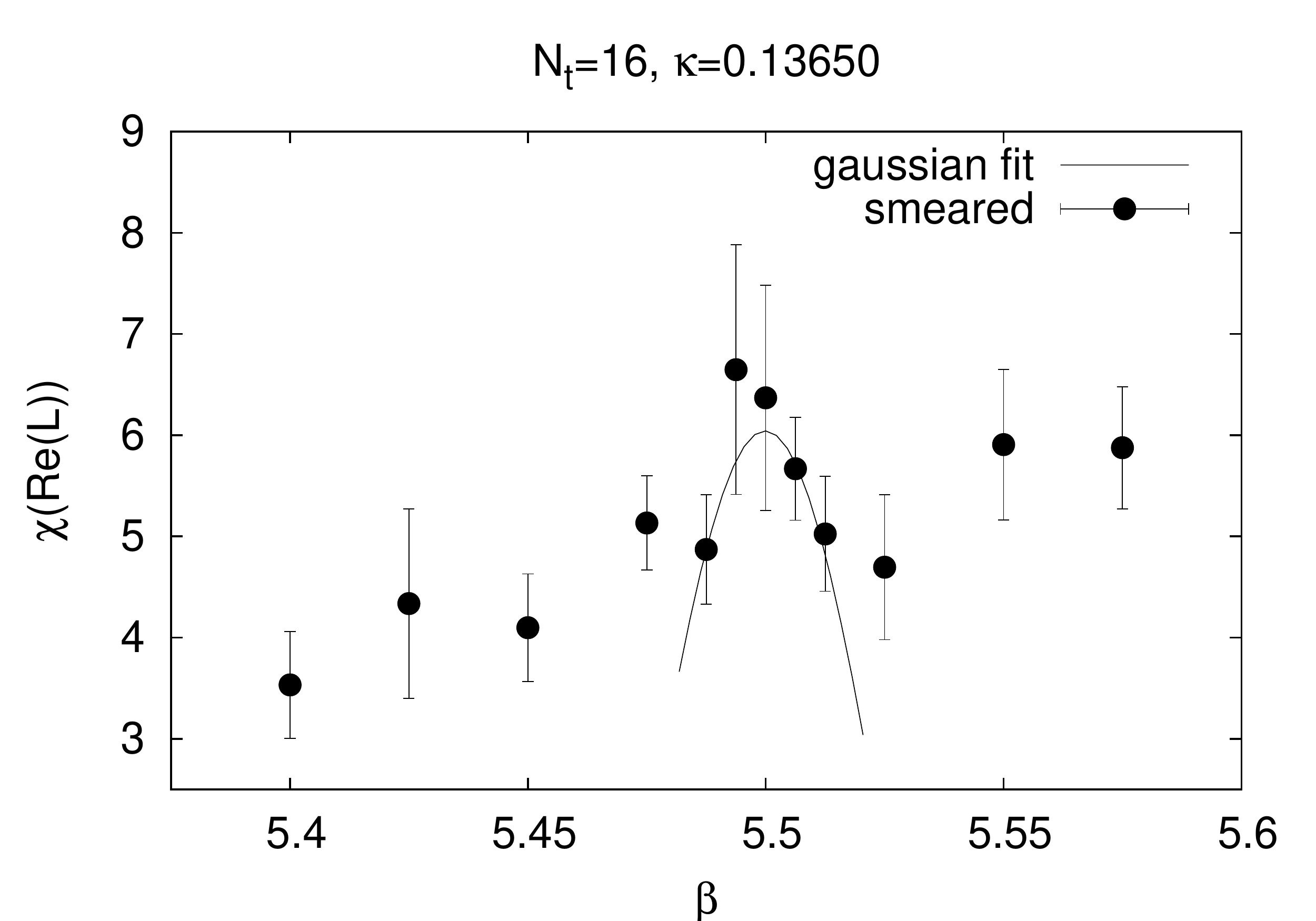}
\end{minipage}
\caption{Plot of the results for scan $B$. {\bf Left:} Real part of the Polyakov loop in the unsmeared and APE-smeared version. The smeared results are rescaled such that we could show them in a single plot. {\bf Right:} The susceptibility of the APE-smeared Polyakov loop, together with a gaussian fit to the points around the peak position.}
\label{fig1}
\end{figure}

\begin{figure}[h]
\centering
\includegraphics[angle=-90, width=.38\textwidth]{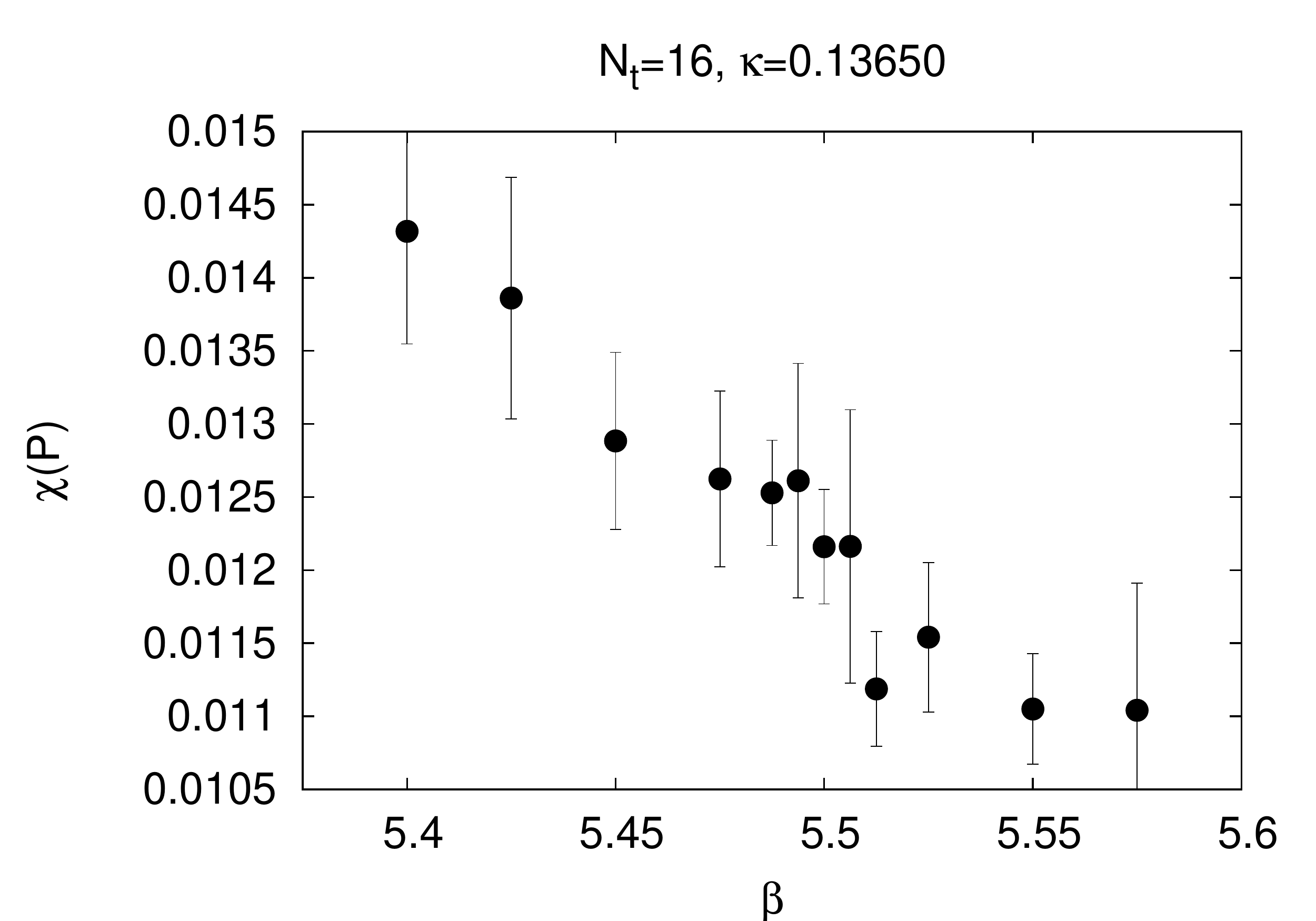}
\caption{The susceptibility of the plaquette from scan $B$.}
\label{fig2}
\end{figure}

So far our simulations where done on two different lattices, with simulation parameters as listed in table \ref{tab1}.

Scan $A$ was designed as a test run for the algorithm and the measurement routines at $T\neq0$. We therefore set the hopping parameter $\kappa$ to the critical value where the phase transition occurs for the real part of the Polyakov loop at the lattice with $\beta=5.29$ of \cite{QCDSF}. We show the signal of the Polyakov loop susceptibility in figure \ref{fig0}. The data is consistent with a phase transition at $\beta_c=5.301(3)$, where we see a strong increase in the signal and a peak in the susceptibility. The resulting transition temperature is slightly higher than the one obtained in \cite{QCDSF}, but mainly consistent when we take into account that at a pion mass of roughly 600 MeV the transition is hardly a sharp phase transition but a broad crossover. For more details see \cite{ownpos}.

Scan $B$ is our first scan at a somewhat lighter pion mass and $N_t=16$. We show the behaviour of $\textnormal{Re}\left[\ev{L}\right]$ together with the smeared version $\textnormal{Re}\left[\ev{L}_{sm}\right]$, and the susceptibility of the latter in figure \ref{fig1}. Compared to \cite{ownpos} we enhanced the resolution around the transition point and increased statistics. At the right of figure \ref{fig1} we also show a Gaussian fit to 5 points around the transition point, from which we obtain the peak position to be $\beta_c=5.499(2)$. The value of $\chi^2/d.o.f.$ is around 1 for the fit. Fortunately there already exists a run for $T=0$ with parameters $\beta=5.50$ and $\kappa=0.13650$ \cite{CLS_POS}, leading to a transition temperature in physical units of roughly $T_c(m_{\pi}=\:510\:\textnormal{MeV},a=0.053\:\textnormal{fm})\approx233$ MeV.
The determination of the scale is still in progress and thus the estimate of $T_c$ in physical units must be considered preliminary. Indeed the new scale determination in \cite{CLS_POS} changed the temperature by around 10\% compared to \cite{ownpos}. Therefore one should keep in mind that the systematic error might still be large. The peak in the susceptibility of the Polyakov loop is reproduced by the other observables as well and we show the behaviour of the susceptibility of the plaquette in figure \ref{fig2}. It is important to note that the susceptibility of the plaquette shows a general decrease due to the behaviour of the corresponding expectation value.

\section{Conclusions and outlook}

In this proceedings article, we give an update of our effort to obtain the QCD deconfinement transition temperature for two dynamical flavours in the chiral limit. Compared to \cite{ownpos}, we have refined the resolution of scan $B$ around the transition point and enlarged statistics. The new scale determination in \cite{CLS_POS} changed the transition temperature for scan $B$ to 233 MeV, which is of the order of the transition temperatures from the twisted mass simulations \cite{tmft} at comparable physical pion mass. The peak in the susceptibilities is reproduced by all observables, as shown for the example of the plaquette in figure \ref{fig2}.

Currently we extend scan $B$ to obtain a finer resolution around the peak and to employ the Multi-Histogram method discussed in \cite{ownpos}. In addition we enlarge the set of scans at $N_t=16$ to lighter pion masses and larger volumes.

\section*{Acknowledgments}

The simulations were done on the WILSON cluster at the University of Mainz (see \cite{CLS}) and on JUGENE at FZ Juelich under NIC Grant Nr. 3330. We are indebted to the institutes for these facilities. We also like to thank H.B. Meyer for many fruitful discussions. B.B. is funded by the DFG via SFB 443. L.Z. is supported by DFG PH 158/3-1.

\end{document}